\documentclass[prl,aps,twocolumn]{revtex4}
\usepackage{graphicx}
\begin{document}

\title{Asymmetric scattering of Dirac electrons and holes in graphene}
\author{Atikur Rahman}
\author{Janice Wynn Guikema}
\author{Nina Markovi\'c}
\affiliation{{\it Department of Physics and Astronomy, Johns Hopkins
University, Baltimore, Maryland 21218, USA.}}

\maketitle

{\bf The relativistic nature of Dirac electrons and holes in graphene
profoundly affects the way they interact with impurities \cite{imp_gr1,
imp_gr2, mc_2, mc_3, mc_4, kotov, rossi}. Signatures of the relativistic behavior have been observed recently in scanning
tunneling measurements on individual impurities \cite{crommie_2}, but the conductance measurements
in this regime are typically dominated by electron and hole puddles \cite{martin, mc_3, mc_4}.
Here we present measurements of quantum interference noise and magnetoresistance in graphene pn junctions.
Unlike the conductance, the quantum interference noise can provide access to the scattering at the Dirac point:
it is sensitive to the motion of a single impurity \cite{ucf_2, rossi, berezovsky}, it depends
strongly on the fundamental symmetries that describe the system \cite{lee_1, bla_1} and it is determined
by the phase-coherent phenomena which are not necessarily obscured by the puddles \cite{rahman}.
The temperature and the carrier density dependence of resistance fluctuations and magnetoresistance in
graphene p-n junctions at low temperatures suggest that the noise is dominated by the quantum interference due
to scattering on impurities and that the noise minimum could be used to determine the point
where the average carrier density is zero. At larger carrier densities, the amplitude of the noise depends strongly
on the sign of the impurity charge, reflecting the fact that the electrons and the holes are scattered
by the impurity potential in an asymmetric manner.}

Conductance fluctuations in disordered electron systems are a consequence of the quantum interference
between the trajectories of  charge carriers scattered on impurities \cite{lee_1, bla_1}. In the case of
static disorder within a phase coherent volume, the conductance
shows reproducible aperiodic fluctuations, known as universal
conductance fluctuations (UCF), as a function of magnetic field or chemical potential
\cite{lee_1}. If the impurity configuration changes slowly over time,
the trajectories of the scattered electrons will also change. This causes the
conductance to fluctuate in time, and the noise power shows a 1/f type
dependence \cite{ucf_2, ucf_4, ucf_5}. One can easily distinguish this mechanism from other sources of 1/f noise
by its temperature dependence. While the conventional noise will typically increase with increasing temperature,
the quantum interference noise will do the opposite: even though the impurities may fluctuate less at lower temperatures,
the phase coherent volume increases, including more impurities and increasing the contribution to the noise.

The effects of quantum interference also lead to corrections to the average conductance: interference between
backscattered paths causes weak localization in disordered electronic systems with weak
spin-orbit coupling \cite{berg_1}. In graphene, the conservation of pseudospin prevents backscattering
on long-range disorder potentials, and localization is expected only due to scattering on atomically sharp,
short-range disorder, which can couple the two valleys \cite{mc_4}.  Quantum corrections and UCF as a function of magnetic field
and chemical potential in graphene have been studied both experimentally and theoretically \cite{mc_4}.

In the vicinity of the Dirac point, however, the predictions of the
quantum transport theory do not strictly hold and analytical
treatment is not always possible. In clean graphene, transport in
this regime is expected to be pseudodiffusive, occurring through
quantum tunneling of evanescent modes \cite{tworzydlo}. In the
presence of disorder, however, the transport is dominated by the
electron and hole puddles \cite{martin,rr_1}. It has also been
argued that charged impurities dominate the conductivity in graphene
and that the minimum conductivity is not determined by the vanishing
carrier density, but rather the added carrier density which
neutralizes the impurity potential \cite{mc_3, adam}.

In this work, we show that the time-dependent resistance fluctuations are determined by a different mechanism than the resistance itself,
and that they are due to fluctuations of the electron trajectories that take part in quantum interference.
We argue that these measurements can access the details of impurity scattering in the vicinity of the Dirac point
and  that the asymmetry in the noise as a function of gate voltage is related to the asymmetry in the scattering of relativistic electrons and
holes by the impurities present in the system.

\begin{figure}
  \includegraphics[width=8cm]{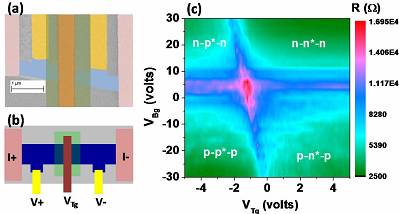}
  \caption{{\bf Sample schematic and resistance as a function of gate voltages.} (a) Colored SEM micrograph of a dual-gated single-layer graphene device. Au leads
were defined on top of a single layer graphene flake using electron beam lithography. Doped silicon substrate is used as a back gate, and the Au top gate is deposited
over an insulating layer of aluminum oxide. The scale bar is 1 $\mu$m. (b) Schematic of the 4-probe measurement configuration for a device shown in a).
  (c) Color plot of the resistance as a function of top gate voltage and the back gate voltage for a dual-gated graphene device (SL1). The carrier type and the carrier density
can be controlled independently by the two gates to create p-p$^*$-p, p-n$^*$-p, n-n$^*$-n and
n-p$^*$-n regions (here $*$ denotes the carrier type under the top gate). For $V_{Tg}=0$, the maximum in the resistance ($V_D$) is located at $V_{Bg}=-6$ volts.
}
  \label{fig1}
\end{figure}

The devices were fabricated by mechanical exfoliation of natural graphite
(see supporting materials for the details of the sample
fabrication). Scanning electron microscope image of a typical device
is shown in Fig. 1(a). All the measurements were done in a 4-probe
geometry (schematic shown in Fig. 1(b)) to minimize the effects of
contacts. In Fig. 1(c) we show the resistance
(measured at 250 mK) as a function of both back gate ($V_{Bg}$) and
top gate voltage ($V_{Tg}$) for a p-n-p transistor device (SL1). All
four regions (marked by p-p$^*$-p, p-n$^*$-p, n-n$^*$-n and
n-p$^*$-n) are clearly visible ($*$ denotes the carrier type under the
top gate). In the absence of top gate voltage, the charge
neutrality point ($V_D$) was found at $V_{Bg}=-6$ volts.

\begin{figure}
  \includegraphics[width=7.5cm]{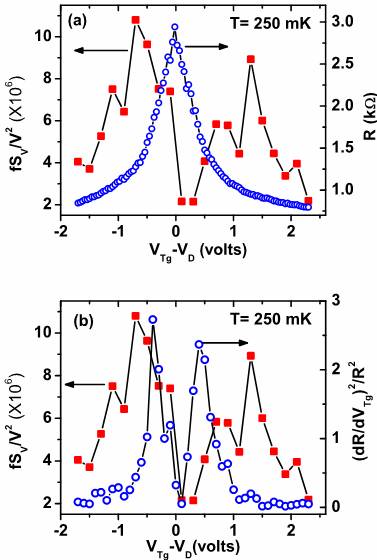}
  \caption{{\bf Noise and the resistance as a function of gate voltage} (a) Normalized noise power (left axis) and resistance (right axis) as a function of the top gate voltage.
The minimum in the normalized noise power does not coincide with the maximum in the resistance.
  (b) $fS_V/V^2$ and $(dR/dV_{Tg})^2/R^2$ as a function of top gate voltage. If the resistance fluctuations were dominated by the fluctuations in
the carrier density, the two curves would be expected to coincide. It is evident that the minimum and the two maxima occur at different values of gate voltage,
The measurements were done at 250 mK.}
  \label{fig2}
\end{figure}

Low-frequency noise measurements were carried out using the ac noise
measurement technique \cite{scofield} (see supporting materials for
the details of the noise measurement). Both invasive and external
voltage probes \cite{external_1} were used for the noise measurements. The power spectral density
($S_V$) shows $1/f^{\alpha}$ dependence at all top gate voltages
with values of $\alpha$ close to 1 (see supporting
materials, Fig. S1). Normalized noise power density ($= fS_V/V^2$ or
$fS_I/I^2$) was found to be independent of the bias current or
voltage, ruling out any effects of heating by the bias
current. The normalized noise power ($fS_V/V^2$) shows a non-monotonic behavior
as a function of top gate voltage. In Fig. 2(a) we show both the
resistance and normalized noise as a function of $V_{Tg}$. It is
important to note that the peak in the resistance and the minimum in the
noise ($S_V$) do not coincide (see supporting materials, Fig.
S2). To ascertain that this is not due to the experimental delay between
the two measurements, we carried out simultaneous measurements of both
noise and resistance as a function of $V_{Tg}$ (see supporting
materials Fig. S2). In several samples, the noise power ($S_V$) itself showed the
non-monotonic behavior as a function of gate voltage
(see supporting materials, Fig. S3
and S4).
The noise measurements were highly reproducible over time at all temperatures,
and did not depend on the direction or scan step of the gate
voltage (see supporting material, Fig. S3 and Fig. S4).

Measurements of 1/f noise in graphene have been
reported before \cite{m1, m2, m3, m4, m5}. The non-monotonic behavior
as a function of gate voltage has been attributed to charge noise
\cite{m1}, interplay between majority and minority carriers
\cite{m2} and competition between long-range and short-range scattering
\cite{m3, m5}. We note that the noise observed here is markedly different from
that discussed above - in our samples, the noise amplitude
\emph {increases} with decreasing temperature, in contrast to the previously reported results
(see supporting material, Fig. S5). Non-monotonic 1/f noise has also been recently predicted
due to non-universal conductance fluctuations in a crossover between a pseudodiffusive and
symplectic regime in clean graphene \cite{rossi}.

In order to explain the origin of the noise, we consider three possible contributions: carrier density fluctuations,
impurity fluctuations and contributions from the pn junction interfaces.

If we assume that the noise is determined entirely by the fluctuations in the carrier density
\cite{number_1} or mobility \cite{mob_1}, then the normalized noise
power, $S_V/V^2$ and the contribution to the resistance fluctuations due to the carrier density $(dR/dV_{Tg})^2/R^2$
should closely follow each other and
exhibit a minimum at the same value of the gate voltage
 \cite{m5,m1}. However, the two do not coincide, as shown in Fig.\ 2(b), indicating that the noise is
not dominated by the fluctuations in the carrier density.

Two main types of impurities are known to play a crucial role in
determining the electronic transport properties of graphene: short-range (SR) and long-
range (LR) impurities \cite{mc_2, mc_3, mc_4}. SR impurities contribute to the resistance
in a density-independent manner \cite{ando_1}, while the effect of LR impurities is
inversely proportional to the density \cite{ando_2}. Consequently,
if scattering on SR impurities were the dominant source of resistance fluctuations, then the normalized noise
would be expected to increase monotonically with increasing carrier concentration ($n$). On the other hand,
if LR impurities dominated the noise, then its magnitude would decrease with increasing $n$. A cross-over
from SR to LR impurity-dominated scattering could therefore account for the nonmonotonic behavior
as a function of gate voltage.

A possible evidence of such a crossover can be found in
the temperature dependence of resistance and
magnetoresistance as a function of $V_{Tg}$, shown in Fig. 3. In Fig. 3(a) we
show $V_{Tg}$ dependence of resistance at two different
temperatures for a typical device (SL3). It is clear that the
resistance increases with decreasing temperature in the vicinity of the Dirac point,
but does not otherwise show a strong temperature dependence.
Normalized noise and the temperature-dependent part of resistance are shown together
as a function of top gate voltage in Fig. 3(b). It is interesting to note that
the regime in which the noise increases as a function of gate voltage
coincides with the range of gate voltages for which the resistance
depends on temperature (shaded region in Fig. 3(b)). Outside the
shaded region, the resistance becomes nearly temperature independent
and the noise starts to decrease as a function of $V_{Tg}$ on both
sides.

\begin{figure}
  \includegraphics[width=8cm]{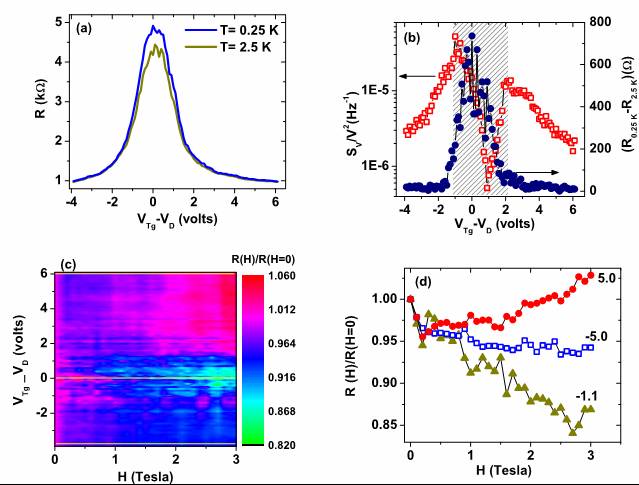}
  \caption{{\bf Temperature and magnetic field dependence of the resistance}(a) Resistance of a typical dual-gated single layer graphene device at two different
  temperatures as a function of top gate voltage. The resistance does not depend on temperature away from
the Dirac point, but increases with decreasing temperature in the vicinity of teh Dirac point. (b) The noise $S_V/V^2$ (right axis) and
R$_{0.25K}$-R$_{2.5K}$ (left axis) as a function of top gate voltage. It is evident that the regime in which the noise increases with
the gate voltage (or carrier density) coinsides with the regime in which the resistance increases with decreasing temperature.
(c) Color plot of the resistance as a function of the top gate voltage and the magnetic field. Negative magnetoresistance is observed
in the vicinity of the Dirac point, with the largest magnitude near the Dirac point ($V_D$=-1.1 V), and decreasing away from it.
In this sample, magnetoresistance remains negative for negative $V_{Tg}-V_{D}$, but it becomes positive at relatively
  large magnetic fields for positive values of $V_{Tg}-V_{D}$. (d) Magnetoresistance as a function of magnetic field for three different values
of top gate voltage (-1.1 V, -5V and 5V). These values are also marked by lines in (c).}
  \label{fig4}
\end{figure}

The increasing resistance with decreasing temperature suggests localization, which should then also be apparent in
magnetoresistance measurements. In Fig. 3(c) and 3(d) we show the magnetoresistance as a function of $V_{Tg}$ for
the same sample. The magnetoresistance is negative and relatively large near the charge neutrality point (CNP).

The negative magnetoresistance might be expected as a result of weak localization
\cite{CG1} and would suggest the important role of short range scatterers  \cite{mc_4}.
Negative magnetoresistance over relatively large magnetic field has also been
observed in chemically modified graphene where short range scatterers
were known to be present  \cite{fl_gr_1, hi_gr_1}.

We find that the sign and the magnitude of the magnetoresistance depends on the position of the CNP
as a function of gate voltage. In particular, if the CNP is
located at negative gate voltage, then we observe
negative MR up to relatively large negative gate voltages. It is evident in Fig.\
3(c) and (d) that the maximum negative MR is observed at
the CNP (located at $V_{Tg}$ = -1.1 volts at zero back
gate voltage). Away from the CNP,  the resistance shows an upturn as a
function of magnetic field on the electron side (positive gate voltage) at a
low gate voltage. On the hole side (negative gate voltage), negative MR is
observed up to large negative gate voltages, though the
magnitude decreases with the increasing gate voltage.
Similarly, if the CNP is located at the positive gate voltage, larger magnetoresistance is observed on the electron side.
Such asymmetric effect of backscattering is consistent with a recent STM observation of quasiparticle interference on localized
scattering centers \cite{crommie_1}. The quasiparticle interference implies backscattering and was observed on scattering centers
only in the electron puddles in the sample in which the CNP was located at positive gate voltage \cite{crommie_1}.

We also find that the position of the CNP affects the noise amplitude away from the CNP.
Specifically, as we move away from the minimum, the noise as a function of gate voltage
becomes asymmetric with respect to the gate voltage, showing two peaks of different heights.
We found that the peak height and position are related to the position of the CNP: if the CNP is
located in the negative side of the gate voltage then the peak in the hole side is larger (Fig.\ 2(a),
3(b), and 5(b)). Similarly, if the CNP is located at a positive value of the gate voltage (electron side), than the noise peak
is larger on the electron side(Fig. 4(d)). We note that an asymmetric contribution to 1/f noise could arise due to
fluctuating gate leakage current \cite{m5}. However, this contribution can be ruled out in our experiment (see supplementary information for details).

\begin{figure}
    \includegraphics[width=8cm]{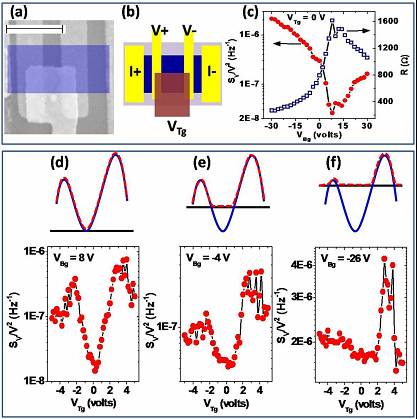}
  \caption{{\bf Parallel pn junction device}(a) Scanning electron microscope image
  of a single-layer parallel p-n junction device (SL4). Position of the graphene flake is indicated by the blue shaded region.
The top gate only partially covers the graphene channel. The scale bar is 1 $\mu$m long.
 (b) Schematic of the four-probe measurement for the parallel p-n junction.
 (c) The noise (left axis) and the resistance (right axix) as a function of back gate voltage for zero
  top gate voltage. The noise minimum is located at $V_{Bg}=8$ V. (d) The schematic and the data of the noise as a function of top gate voltage at $V_{Bg}=8$ V.
The blue solid line represents the noise signal from the portion of the sample under the top gate, the straight black line represents the noise signal from the portion
of the sample that is not covered by the top gate (this part is independent of the top gate voltage), and the red dashed line indicates the total noise signal that would
be measured, to be compared with the data below the schematic. Similar schematics and the data are shown in (e) for $V_{Bg}=-4$ V and in (f) for $V_{Bg}=-26$ V.}
  \label{fig5}
\end{figure}

To rule out the possibility that the measured noise is generated at the p-n junction interfaces
(between two regions characterized by two different types of charge carriers), we fabricated
a parallel p-n junction \cite{marcus_1} (shown in Fig.\ 4(a)) and
studied the noise characteristics as a function of $V_{Bg}$ and
$V_{Tg}$. As the top gate voltage affects half of the sample in a
parallel configuration (see Fig.\ 4(b) for a schematic), the
resistance of the sample will be determined by the parallel
combination of the resistance of the two segments (portions with and
without top gate) and lowest resistive path will dominate. However,
the noise of the entire system will be determined by the maximum noise in either of the
two segments. In the absence of the top gate voltage, the minimum in the noise is
found at $V_{Bg} = +8$ volts (Fig.\ 4(c)). As a function of $V_{Tg}$, two maxima appear in the noise (Fig.\ 4(d)).
As the back gate is set at the value at which the noise shows a minimum, we can safely assume
that the maxima in the noise arise from the part of the sample
under the top gate. We found that the peak on the electron side is
larger than the peak on the hole side in this case (Fig.\ 4(d)). Away from
the CNP the noise increases with  $V_{Bg}$ on both sides (Fig.\ 4(c)).
As a result, the shape of
the $S_{V}/V^2$ vs. $V_{Tg}$ characteristics changes continuously
with changing $V_{Bg}$ (Fig.\ 4(e) and (f)). It is evident from the
data that noise is always large when the carriers under the top gate
are electrons. In Fig.\ 4(f) we see that for $V_{Tg}$ = 0
volts the carriers are holes ($V_{Bg}$ = -26 volts). If we
add more holes by applying a negative $V_{Tg}$, we find that the noise
increases slightly. On the other hand, for positive $V_{Tg}$, the charge carriers
under the top gate are electrons and a sharp increase in noise is
observed. This excess noise on the electron side suggests that the scattering by the impurities present
in the system has a stronger effect on electrons compared to holes.

\begin{figure}
    \includegraphics[width=8cm]{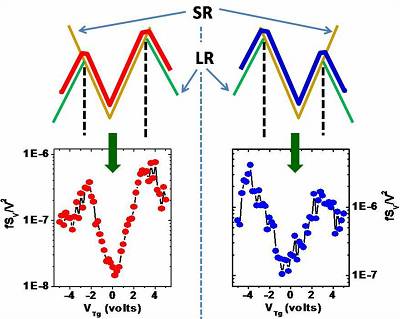}
  \caption{{\bf The origin of the asymmetric shape} A model and and the data showing how the asymmetric shape develops. In the
  upper left and right panel, the yellow curve indicates dependence of noise power on the gate voltage
  for short range scatterers. The vertical black dashed lines indicate the extent of the gate voltage up to which the short-range
  scatterers dominate. At larger values of gate voltage (corresponding to larger carrier densities), the long range scatterers dominate the noise and the noise
  decreases with further increase of gate voltage (denoted by the green line).
Depending on the position of the CNP, the effect of the backscattering (denoted by the black vertical dashed line) is extended more towards either
  electron side (for positive CNP) or hole side (for negative CNP). Representative experimental data are shown for the cases in which the CNP was located
  at +8V (lower left panel, device SL4) and -10V (lower right panel, device SL6).}
  \label{fig6}
\end{figure}

The observed noise is clearly not due to the noise generated at the p-n
junction interface, because then we would not observe it in case of a
parallel p-n junction. This also implies that the electron and hole puddles that
form near the CNP are not responsible for the observed noise.
Specifically, if the noise is determined by the resistance fluctuations
at the p-n junctions formed by the puddles, then we would expect the noise to
decrease with increasing gate voltage away from the CNP.
As we move away from the CNP by increasing the gate voltage, the typical size of one type of puddles will
decrease, so the effective p-n junction area will also decrease. Consequently, any noise
generated by the current floeing through a pn junction interface should decrease with increasing gate voltage, opposite
to our observations close to CNP.
In the random resistor model of puddles in graphene, the resistance is
determined by the weak links between p and n type regions \cite{rr_1}. Any fluctuation of the junction
resistance will give rise to low frequency noise. However, if the noise were
generated due to the fluctuations of the p-n junction resistance,  then
we would not expect to see a strong temperature dependence, since the
resistance in graphene does not vary strongly with temperature (see supporting
materials Fig. S5).

We also found that the negative magnetoresistance persists up to
relatively large value of negative gate voltage (hole-type carriers),
when the CNP is located at the negative side of the gate voltage and
vice versa. We note that magnetoresistance itself could possibly be due to the transmission through the pn interfaces
formed by electron and hole puddles at low carrier densities, although it is not obvious that this mechanism would result
in negative magnetoresistance \cite{falko}.

It is evident that the resistance increases with decreasing temperature in the same range of gate voltages
for which the noise increases as a function of gate voltage (shaded region in Fig. 3.b). The maximal change in the resistance
does not coincide with the minimum in the noise, which suggests that the resistance and the noise
are dominated by different phenomena. This is also the same regime in which we see the largest negative magnetoresistance.
All three phenomena (increasing resistance with decreasing temperature, increasing noise as a function of gate voltage and
negative magnetoresistance) could be explained in terms of scattering on SR impurities.
STM measurements support the relevance of SR scatterers near the Dirac point: the results on epitaxialy grown
graphene found that SR scatterers are the dominant scattering centers \cite{rutter}.  A more recent STM experiment
found scatterers of the SR type only in the electron puddles in samples in which the charge neutrality point was located on
the electron-doped side \cite{crommie_1}, which is consistent with the asymmetry observed in our experiments.

Keeping this in mind, we can understand the asymmetric shape
of the noise through a model shown in Fig.\ 5. For a positive value of
CNP,  the backscattering and the increase of noise is observed over a large positive gate voltage.  Assuming that the SR scatterers
dominate in this regime, we would expect the noise to increase with increasing carrier density.
This also implies that it is actually the \emph{minimum in the noise}, not the maximum in resistance, that reflects the vanishing carrier density.
If we increase the gate voltage further, the noise becomes dominated by LR scatterers
and we expect to see a decrease in the noise with increasing gate voltage.
Taken together, these two effects result in a non-monotonic and asymmetric shape of the noise as a function of the gate voltage.
Similar results are obtained in the case when the CNP is located at a negative value of the gate voltage:
 Fig.\ 5 shows the model and the corresponding data for CNP
located at +8 volts (left panel) and -10 volts (right panel).

We note that a qualitatively similar non-monotonic behavior in the quantum interference noise was predicted recently due to
a crossover from the pseudodiffusive to symplectic metal regime \cite{rossi}. While this could easily describe our noise data,
it is not obvious that our samples are in the relevant regime. More detailed theoretical treatment would be needed to show whether
the temperature dependence and magnetoresistance observed in our experiments are also consistent with such a crossover.

Unlike conductivity, which is dominated by charge impurities \cite{mc_3,adam}, we find that the quantum interference noise is dominated by
scattering on disorder of a short-range type in the vicinity of the Dirac point. The observed asymmetry in the noise with respect to the gate voltage
suggests that electrons and holes scatter differently from the disorder potential. The nature of this asymmetry is consistent with that
observed in conductivity \cite{mc_3} and STM  \cite{crommie_1} measurements, and can be related to the relativistic nature of electrons
and holes in graphene. The minimum in the noise can be used to determine the gate voltage at which the carrier density vanishes,
which, in disordered graphene, is not identical to the point at which the conductivity shows a minimum. This shows that quantum interference noise
can be used as a probe that is complementary to conductivity and STM measurements and can be used to investigate the details of scattering of
relativistic Dirac fermions.

\end{document}